\documentclass[aip, apl, amsmath, numerical, amssymb, reprint]{revtex4-1}

\usepackage{graphicx}% Include figure files
\pdfoutput=1
\usepackage{dcolumn}% Align table columns on decimal point
\usepackage{bm}% bold math
\usepackage{epstopdf} % Fixes compile error that file figure is not found.
\usepackage[utf8]{inputenc}
\usepackage[T1]{fontenc}
\usepackage{mathptmx}
\usepackage[dvipsnames]{xcolor}

\begin{document}

	\title{Thermal stability for domain wall mediated magnetization reversal in perpendicular STT MRAM cells with W insertion layers}
	% Force line breaks with \\
	
	\author{G.\ Mihajlovi\'{c}}
	\email{goran.mihajlovic@wdc.com}
	\affiliation{Western Digital Research Center, Western Digital Corporation, San Jose, CA 95119\\}
	
	\author{N.\ Smith}
	\email{neil.smith@wdc.com}
	\affiliation{Western Digital Research Center, Western Digital Corporation, San Jose, CA 95119\\}
	
	\author{T.\ Santos}
	%\email{tiffany.santos@wdc.com}
	\affiliation{Western Digital Research Center, Western Digital Corporation, San Jose, CA 95119\\}
	
	\author{J.\ Li}
	%\email{goran.mihajlovic@wdc.com}
	\affiliation{Western Digital Research Center, Western Digital Corporation, San Jose, CA 95119\\}
	
	\author{B.\ D.\ Terris}
	%\email{goran.mihajlovic@wdc.com}
	\affiliation{Western Digital Research Center, Western Digital Corporation, San Jose, CA 95119\\}
	
	\author{J.\ A.\ Katine}
	%\email{goran.mihajlovic@wdc.com}
	\affiliation{Western Digital Research Center, Western Digital Corporation, San Jose, CA 95119\\}
	
	\date{\today}
	
\begin{abstract}
We present an analytical model for calculating energy barrier for the magnetic field-driven domain wall-mediated magnetization reversal of a magneto-resistive random access memory (MRAM) cell and apply it to study thermal stability factor $\Delta$ for various thicknesses of W layers inserted into the free layer (FL) as a function of the cell size and temperature. We find that,  by increasing W thickness, the effective perpendicular magnetic anisotropy (PMA) energy density of the FL film monotonically increases, but at the same time, $\Delta$ of the cell mainly decreases. Our analysis shows that, in addition to saturation magnetization $M_s$ and exchange stiffness constant $A_\mathrm{ex}$ of the FL film, the parameter that quantifies the $\Delta$ of the cell is its coercive field $H_c$, rather than the net PMA  field $H_k$ of the FL film comprising the cell.
\end{abstract}
	
\maketitle
	
Thermal stability factor $\Delta$ quantifies retention of the spin-transfer-torque magneto-resistive random access memory (STT MRAM) cell. It is defined as the ratio of the energy barrier $E_b$ for magnetization $M$ reversal of the free layer (FL) of the magnetic tunnel junction (MTJ) comprising the memory cell, and the thermal energy $k_B T$, i.e. $\Delta =  E_b / (k_B T)$ ($k_B$ is the Boltzmann constant and $T$ is temperature). For a required small memory chip bit error rate $\mathrm{BER} \ll 1$ against thermal bit flip \cite{Brown_PR1963, Meo_SciRep2017} and a retention time $t$, $\Delta_{\mathrm{eff}} > \ln(f_0 t/\mathrm{BER})$  is required, where $f_0$ = 1~GHz is the  attempt frequency  and $\Delta_{\mathrm{eff}}  = \Delta_m- \sigma^2 / 2$ is the effective $\Delta$ value for the memory chip \cite{Thomas_APL2015}. The latter expression assumes normal distribution of  $\Delta$ values of individual chip cells, with $\Delta_m$ and $\sigma$ being the median and the standard deviation of the distribution. While the direct way to evaluate $\Delta_{\mathrm{eff}}$ is by examining fraction of bits flipping their $M$s as a function of $t$ at various $T$s (the so-called retention bake method)\cite{Thomas_APL2015}, this approach is rarely used as it is time consuming. Among many various alternative techniques \cite{Tillie_IEDM2016}, Thomas et al. \cite{Thomas_IEDM2015} showed that, for the MTJ diameters  $D > 55$~nm,  $\Delta_m$ and $\sigma$ values can also be determined by fitting the magnetic field $H$ induced empirical switching probabilitiy distributions $P(H)$ of individual cells to a Neel-Brown relaxation model \cite{Brown_PR1963}, namely

\begin{equation}
P(H) = 1 - f_0 t \exp\left[-\Delta \left(H \right) \right] \equiv 1 - f_0 t \exp\left[-\frac{E_b \left(H \right)}{k_B T} \right],
\label{eq:1}
\end{equation}
with $E_b (H)$ calculated assuming domain wall-mediated $M$ reversal (DWMR). Micro-magnetic \cite{Yoshida_JJAP2019} and atomistic \cite{Meo_arXiv2019} simulations suggest that MTJ FL with perpendicular magnetic anisotropy (PMA) can prefer DWMR down to $D \approx 25$~nm. In that case, $E_{b, \mathrm{dw}} = D \varepsilon_{\mathrm{dw}} t_{\mathrm{FL}}$ where $t_{\mathrm{FL}}$ is the FL thickness and $\varepsilon_{\mathrm{dw}} = \sqrt{8 M_s H_k A_{\mathrm{ex}}}$ is the DW energy density ($M_s$ is  saturation magnetization, $H_k$ is the net PMA field and  $A_{\mathrm{ex}}$ is the exchange stiffness constant of the FL). Alternatively, $\varepsilon_{\mathrm{dw}} = 4 \sqrt{ K_{\mathrm{eff}} A_{\mathrm{ex}}}$ where $K_{\mathrm{eff}} = M_s H_k / 2$ is the FL PMA volume energy density. Thus, for the DWMR, $\Delta$ is expected to depend on  $A_{\mathrm{ex}}$ and  $K_{\mathrm{eff}}$ (i.e. $M_s$ and $H_k$) of the FL film. In addition to sandwiching the CoFeB-based FL between two MgO layers \cite{Sato_APL2012, Jan_APEX2012}, a common approach to increasing $H_k$ is by inserting a thin non-magnetic layer, typically Ta \cite{Sato_APL2012, Couet_APL2017}, Mo \cite{Almasi_APL2015, Almasi_APL2016} or W \cite{ Kim_SciRep2015, Couet_APL2017}. Recently, however, it has been reported that such insertion layers (ILs) dilute magnetic moment of the FL which reduces $M_s$, resulting in poor thermal stability performance at higher operational $T$s \cite{IwataHarms_SciRep2018}. Another study found that W ILs result in reduction of zero-$T$ $A_{\mathrm{ex}}$, suggesting potential disadvantage of such ILs for achieving high $\Delta$ of MRAM cells for DWMR \cite{Mohammadi_ACSAEM2019}. Ultrathin CoFeB FLs that provide superior STT switching performance without utilizing heavy-metal ILs to promote PMA \cite{IwataHarms_SciRep2019} have also been reported. In all these studies, $\Delta$ values for MRAM cell for DWMR were estimated using  $H_k$ values measured on the full FL film. However, a direct, device-level study of the physical parameters quantifying $\Delta$ for DWMR of the FLs at operation-relevant $T$s has been lacking. 

Here we present an experimental study of $P(H)$ for perpendicular MRAM cells fabricated from FL films having variable thickness of the  W IL $t_{\mathrm{W}}$ at $T =$~30, 85, and 125~$^\mathrm{o}$C. We describe an analytical model for calculating $E_{b, \mathrm{dw}}$ which introduces correction to the droplet model  previously used in literature \cite{Hinzke_PRB1998, Chaves-OFlynn_PRA2015, Thomas_IEDM2015, Meo_arXiv2019}. By fitting $P(H)$ using this model, we find that $\Delta$ of MRAM cells decreases with increasing $t_{\mathrm{W}}$, even though  $H_k$ and  PMA energy per unit area $K_{\mathrm{eff}} t_{\mathrm{FL}}$ of the full FL film are increasing. We show that this is not only due to decreasing $M_s$ and $A_{\mathrm{ex}}$ with increasing  $t_{\mathrm{W}}$, but also due to decreasing $H_c$ of the cell. We determine DW width $w_{\mathrm{dw}}$ in range of 11 - 17~nm  for $T$s in range 30 - 125~$^\mathrm{o}$C, with largest values at highest $T$. Our results and analysis provide valuable insights into physical factors important for achieving high MRAM cell $\Delta$ for technologically  relevant cell sizes and operational $T$s.

In the basic droplet model \cite{Meo_arXiv2019, Hinzke_PRB1998, Chaves-OFlynn_PRA2015}, a zero-width domain wall inside a circular FL is taken to be a circular arc of radius $r$ forming a right-angle with the FL perimeter (see Fig.~1(a)). This model, however, predicts that $E_{b, \mathrm{dw}}(H) > 0$ for all finite $H$, an unphysical feature that is removed by including a finite domain wall width $w_{\mathrm{dw}}$ \cite{Thomas_IEDM2015}. The latter can be implemented by reducing the area of both domains via modulation of the wall position (see Fig.~1(b)). In Ref.~[5], this was done by modulating length of the radial coordinate as $\Delta r \leftrightarrow \pm w_{\mathrm{dw}} / 2$, while maintaining the right-angle constraint. However, this approach is mathematically flawed, since  position of the right/exterior end of the radial line $r$ (i.e. point O in Figs.~1(a-b)) will also be modulated. From Fig.~1, the correct identification is $\Delta x \leftrightarrow \pm w_{\mathrm{dw}} / 2$, with the right end of line $x$ fixed in position at the perimeter (point P in Figs.~1(a-b)). A more complete description of the required mathematical transformations can be found in Section I of the Supplementary Material. Defining $q \equiv x / R$,  $\delta \equiv w_{\mathrm{dw}} / D$ and $q_{\pm} \equiv q \pm \delta$, the results are: 

\begin{figure}
	\includegraphics [scale = 0.4, bb = 0 0 590 220]{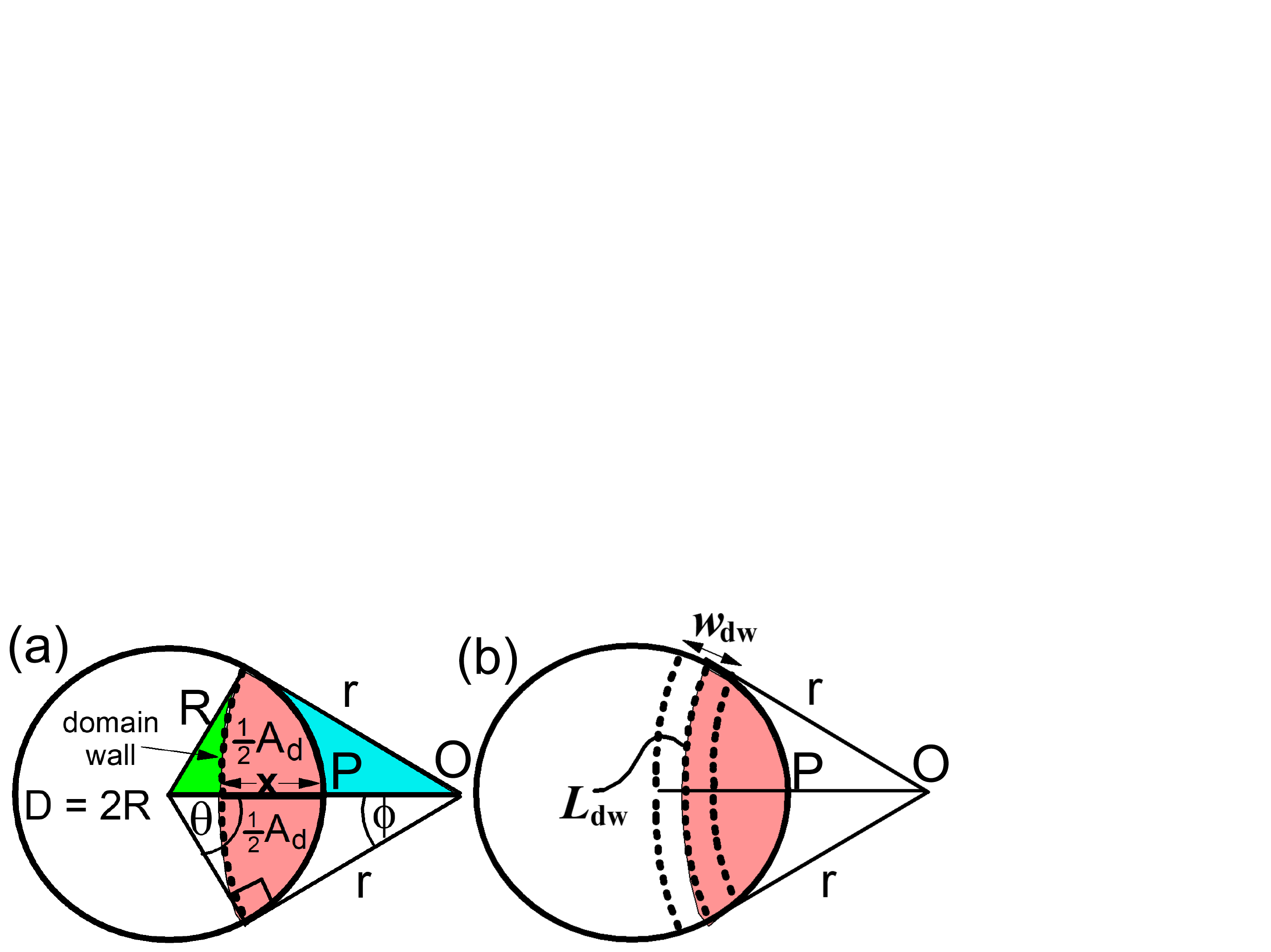}
	\caption{Illustrations of (a) circular FL with reversed domain area $A_d$ formed at a distance $x$ from the right edge and (b) DW length $L_{\mathrm{dw}}$ and finite DW width $w_{\mathrm{dw}}$.}
	\label{fig1:DW reversal illustration}
\end{figure}

\begin{subequations}
	\begin{align}
		\theta (q) = \tan^{-1} \left[\frac{q \left( 1 - q / 2\right) }{1 - q}\right], \\	
		A_d(\theta) = \frac{D^2}{4} \left[\theta - \tan \theta + \left(\frac{\pi}{2} - \theta \right) \tan^2 \theta  \right], \\		
		L_{\mathrm{dw}}(\theta) = D \left(\frac{\pi}{2} - \theta \right)  \tan \theta, \\		
		E(q) = L_{\mathrm{dw}}(q) \varepsilon_{\mathrm{dw}} t_{\mathrm{FL}}  + |H| M_s    t_{\mathrm{FL}} \left[ \frac{D^2 \pi}{4} - A_d(q_+)  - A_d(q_-)\right], \\		
		E_{b, \mathrm{dw}}(H; \delta) \cong E(q_1) - \frac{\pi}{4} D^2 M_s H t_{\mathrm{FL}}, \\		
		q_1 = 1 + \epsilon - \sqrt{1 + \epsilon^2}, \\
		\epsilon = \frac{\varepsilon_{\mathrm{dw}}}{M_s \left| H \right| D }.
	\end{align}
\end{subequations}

The solutions (2e)-(2g) for $E_{b, \mathrm{dw}}(H)$ are accurate to first order in $\delta$ and can be used to determine $\varepsilon_{\mathrm{dw}}$ and $w_{\mathrm{dw}}$ by fitting $P(H)$ using Eq.~(1). A 2nd order solution with $q_1 \rightarrow q_1 + \beta \delta^2$ is described in the Supplemental Material, though the difference is quite small in practical cases. By contrast, the $r$-based solution of Ref.~[5] has leading error term of order $\delta$. 

The MRAM film stacks used in this study consist of a Ta/Pt seed layer (8.0 nm), (Co/Pt)/Ru/(Co/Pt)/CoFeB synthetic antiferromagnet reference layer (RL) (7.1 nm), MgO tunnel barrier, CoFeB/CoFe/W/CoFe FL, MgO cap layer for enhancing $H_k$, and Ru/Ta cap layer (3 nm). In the FL, the thickness of the magnetic layers was fixed at 1.5~nm, while the thickness of the W spacer layer was varied, i.e. $t_{\mathrm{W}} =$~1.1, 1.5, 2.0 and 2.6~\AA{}. The films were deposited by magnetron sputtering in an Anelva C-7100 system and then annealed at 335$^\circ$C for 1 hour. The MgO layers were rf-sputtered from a MgO target. $\mathrm{RA} \cong 11$~$\Omega \mu$m$^2$ and $\mathrm{TMR} \cong 140 \%$ values are measured on the annealed films by current-in-plane tunneling (CIPT) \cite{Worledge&Trouilloud_APL2003} at room $T$ (see Table~I).  $M_s$ was obtained by measuring the magnetic moment of the FL using vibrating sample magnetometry and dividing it with the FL volume assuming the full FL thickness $t_{\mathrm{FL}}$, including $t_{\mathrm{W}}$.  $H_k$ of the FL was measured by full film ferromagnetic resonance. From Table~I, one can see that while $M_s$ monotonically decreases with increasing $t_W$, $H_k$ and $K_{\mathrm{eff}} t_{\mathrm{FL}}$ calculated using corresponding $M_s$ and $H_k$ values increase. However, $K_i$ decreases with increasing $t_{\mathrm{W}}$ (see the last column of Table~I), implying that intrinsic interfacial PMA also decreases with increasing $t_{\mathrm{W}}$ and that the net increase in $H_k$  and $K_{\mathrm{eff}} t_{\mathrm{FL}}$ at room $T$ is solely due to reduction of the $M_s$ of the FL.

\begin{table}
	\caption{\label{tab:table1} Transport and magnetic properties of free layer films used in this study. The parameter values are expressed to the last significant digit based on the corresponding error analysis. $K_i$ is the intrinsic interfacial PMA surface energy density, physically independent of $M_s$, but here extracted from the values of $M_s$ and $H_k$ using the relation  $K_i = M_s t_{\mathrm{FL}} (H_k + 4 \pi M_s) / 2$.}
	\begin{ruledtabular}
		\begin{tabular}
			{ccccccccc}& $t_\mathrm{W}$ & RA        & TMR    & $M_\mathrm{s}$            & $H_\mathrm{k}$  & $K_{\mathrm{eff}} t_{\mathrm{FL}}$ & $K_i$\\
			&(\AA{}) & ($\Omega \mu$m$^{\mathrm{2}}$)  & (\%)     & (emu/cm$^\mathrm{3}$)      & (kOe)  & (erg/cm$^\mathrm{2}$)  & (erg/cm$^\mathrm{2}$)      \\
			\hline
			&1.1  & 11.2                     & 135      & 1495              & 1.81   & 0.22 & 2.48  \\
			&1.5  & 11.1                     & 142      & 1360              & 3.03   & 0.34 & 2.26 \\
			&2.0  & 11.1                     & 141      & 1211              & 4.25   & 0.44 & 2.00   \\
			&2.6  & 11.2                     & 134      & 1052              & 5.14   & 0.48 & 1.70  \\
		\end{tabular}
	\end{ruledtabular}
\end{table}

Circular MRAM test device cells are fabricated using 193~nm deep UV optical lithography, followed by reactive ion etching a hard mask, ion milling the MRAM film, SiO$_2$ refill and chemical mechanical planarization. Electrical critical dimension CD of each device is determined using film-level RA provided in Table~I and the device resistance in parallel state $R_\mathrm{P}$ as $\mathrm{CD} = \sqrt{4 \mathrm{RA} / (\pi R_\mathrm{P})}$. CD values were clustered around four target sizes of 65, 90, 105 and 120~nm.  

The $P(H)$ distributions are obtained by sweeping $H$ via a staircase ramp with the step of 5~Oe and with a dwell time of 0.2 ms, and measuring the MTJ resistance $R$ at low bias voltage of 10~mV to minimize STT effects on $M$ reversal. Fig.~2(a) shows an example of one hundred $R$~vs~$H$ transfer loops measured at $T = 30$~$^\mathrm{o}$C, for a device having $\mathrm{CD} \equiv D \cong 65$~nm, while the corresponding empirical $P(H)$ and fit to Eq~(1) using $E_b (H)$ as expressed in Eqs.~(2e)-(2g) is shown in Fig.~2(b). $\mathrm{P} \rightarrow \mathrm{AP}$ and $\mathrm{AP} \rightarrow \mathrm{P}$ branches are fit simultaneously by substituting $H \rightarrow H - H_{\mathrm{offset}}$ with fixed $H_{\mathrm{offset}} = H_{\mathrm{P} \rightarrow \mathrm{AP}}^{P = 0.5} - H_{\mathrm{AP} \rightarrow \mathrm{P}}^{P = 0.5}$ and fixed $M_s$, leaving  $\varepsilon_{\mathrm{dw}}$ and $\delta$ as the only two fitting parameters. From here, $\Delta = D t_{\mathrm{FL}} \varepsilon_{\mathrm{dw}}/k_B T$, $w_{\mathrm{dw}} = D \delta$. For the device shown in Fig.~2(b) we obtained $w_{\mathrm{dw}} = 12.7$~nm and $\varepsilon_{\mathrm{dw}} = 6.2$~erg/cm$^2$, corresponding to $\Delta = 154$.

\begin{figure}
	\includegraphics [scale = 0.5, bb = 0 0 400 500]{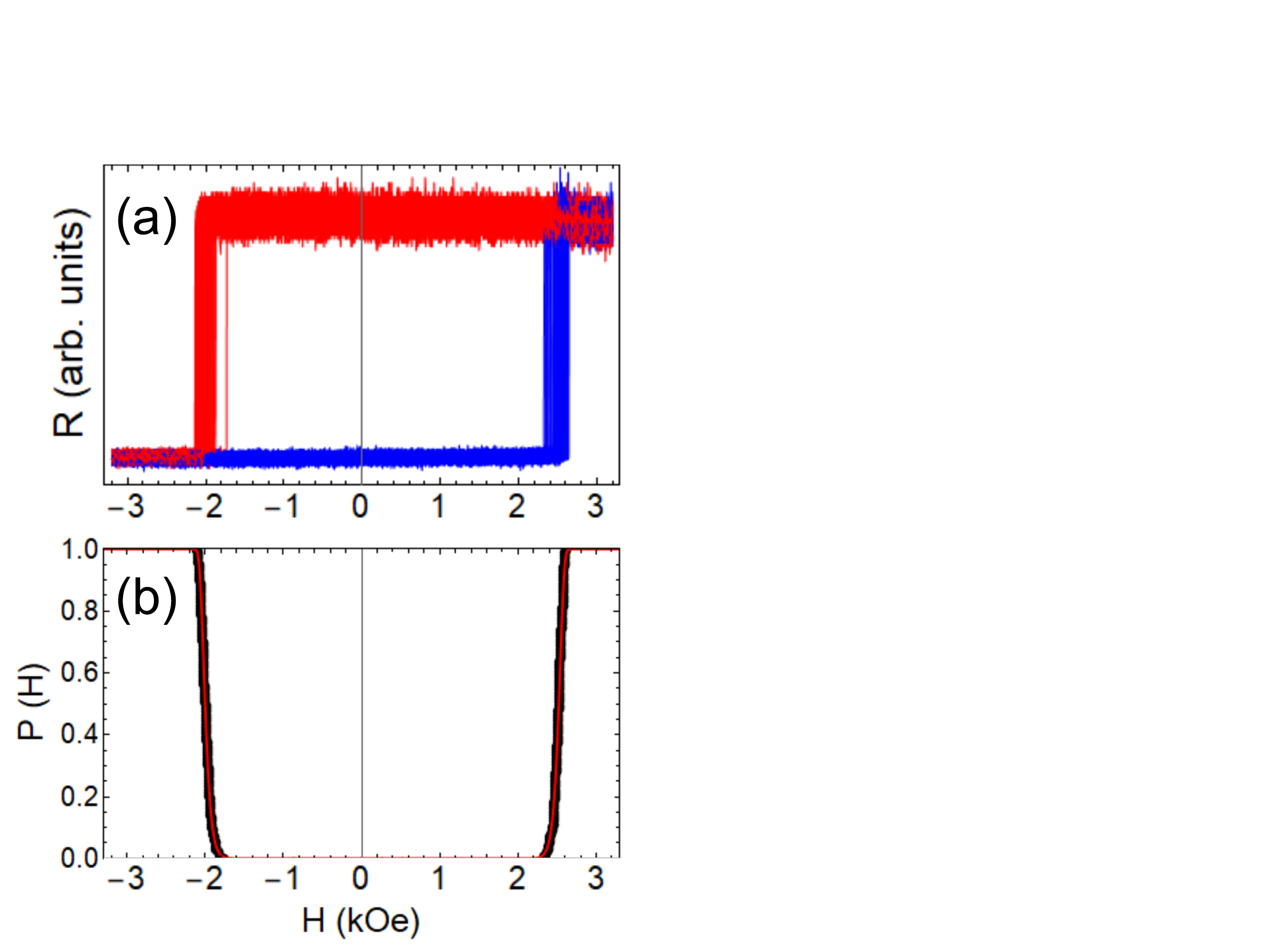}
	\caption{(a) One hundred $R$~vs~$H$ loops measured for a MRAM cell with $t_\mathrm{W} = 1.1$~\AA{}  and $D = 65$~nm at $T = 30$~$^\mathrm{o}$C. (b) Empirical $P(H)$ (black circles) corresponding to switching fields shown in part (a) of the Figure and fit to DWMR model (red lines) as described in text.}
	\label{fig2:P(H) example}
\end{figure}

Fig.~3(a) shows $\varepsilon_{\mathrm{dw}}$, obtained by fitting $P(H)$ using Eqs.~(1)-(2), as a function of $D$ at $T = 30$~$^\mathrm{o}$C for FLs with different $t_\mathrm{W}$.  $\varepsilon_{\mathrm{dw}}$ increases for thinner W, as well as for smaller $D$. The latter is quite moderate and can be attributed to mildly increasing $H_k$ with decreasing $D$ due to reduction of dipolar shape anisotropy.  On the other hand, the dependence on $t_\mathrm{W}$ is much stronger, resulting in about 30 \% increase of $\varepsilon_\mathrm{dw}$, from approximately 4.5~erg/cm$^2$ for $t_\mathrm{W} = 2.6$~\AA{} to approximately 6~erg/cm$^2$ for $t_\mathrm{W} = 1.1$~\AA{}. This suggests that reduction of $H_k$ with decreasing $t_\mathrm{W}$ is more than compensated by increases in $M_s$ and, possibly, $A_\mathrm{ex}$. 

Fig.~3(b) shows the corresponding $\Delta$ values as a function of $D$. $\Delta$ is highest for the FL with lowest $t_\mathrm{W}$,  and lowest for the FL with highest $t_\mathrm{W}$. This is in direct opposition to the trends of $H_k$ and $K_{\mathrm{eff}} t_{\mathrm{FL}}$ (see Table~I). One can also see that while the difference in $\Delta$ values between thinnest and thickest W IL is significant, this difference is quite small between FLs with  1.1~\AA{} and 1.5~\AA{} W IL. This suggests that for the given magnetic thickness and composition of the FL in our study, the optimal $t_\mathrm{W}$ resulting in maximum $\Delta$ is $\approx 1$~\AA{}.

Another parameter obtained directly by fitting $P(H)$ is $w_{\mathrm{dw}}$, shown in Fig.~3(c). We find  $w_{\mathrm{dw}} \cong 11 - 15$~nm for all $D$ and  $t_\mathrm{W}$ values. These are smaller than reported previously for PMA thin films where $w_{\mathrm{dw}}$ was measured by magneto-optical Kerr microscopy imaging\cite{Yamanouchi_IEEEML2011, Buford_IEEEML2016a}, but are consistent with device-level micromagnetic simulation results \cite{Yoshida_JJAP2019}.  However, unlike   $\varepsilon_{\mathrm{dw}}$ and $\Delta$, $w_{\mathrm{dw}}$ does not depend monotonically on $t_\mathrm{W}$: it is largest for $t_\mathrm{W} = 1.1$~\AA{}, smallest for $t_\mathrm{W} = 1.5$~\AA{}, while  the values for $t_\mathrm{W} = 2.0$~\AA{} and 2.6~\AA{} are in between. 

The co-existence of highest $\varepsilon_{dw}$ and $w_{\mathrm{dw}}$ for thinnest W, suggests largest $A_{ex}$ for this case, as both are $\varpropto \sqrt{A_{ex}}$ ($w_\mathrm{dw} = 2 \ln 2 \sqrt{A_\mathrm{ex} / K_\mathrm{eff}}$, see Section II of Supplementary Material for derivation of this expression). This is indeed the case, as can be seen  in Fig.~3(d) where we plot calculated $A_{ex} =  \varepsilon_\mathrm{dw} w_\mathrm{dw} / (8 \ln 2)$. Indeed, $A_\mathrm{ex}$ is highest for thinnest W and decreases with increasing W thickness. This decrease, averaged over all device sizes, is monotonic (see also Fig.4(e)). This is consistent with previously reported findings \cite{Buford_IEEEML2016a, Mohammadi_ACSAEM2019}, although the values $A_{ex} \cong 1.0 - 1.7$~$\mu$erg/cm that we obtain are approximately a factor of 2-3 higher compared to these reports. Thus, based on simple relation between  $w_\mathrm{dw}$, $A_\mathrm{ex}$ and $K_\mathrm{eff}$, the non-monotonic dependence of $w_{\mathrm{dw}}$ on $t_\mathrm{W}$ suggests non-monotonic dependence of $K_\mathrm{eff}$ and likely $K_\mathrm{eff} t_\mathrm{FL}$ in our fabricated cells, contrary to the results obtained from full films (see Table I). This is indeed the case, as can be seen in Fig.~3(f) where we plot $K_\mathrm{eff} t_\mathrm{FL} = (\ln2 / 2) (\varepsilon_\mathrm{dw} / w_\mathrm{dw}) t_\mathrm{FL}$ (full symbols). $K_\mathrm{eff} t_\mathrm{FL}$ increases when $t_\mathrm{W}$ is increased from  1.1~\AA{} to 1.5~\AA{}, but it then decreases for larger $t_\mathrm{W}$. Surprisingly, the observed trend can be well reproduced by assuming that $K_\mathrm{eff}$ in our devices is determined by device-level coercive field $H_c$ (see Fig.~3(e)) instead of the film level $H_k$, i.e. $K_\mathrm{eff} = M_s H_c / 2$. The $K_\mathrm{eff} t_\mathrm{FL}$ values calculated this way are shown in Fig.~3(f) as open symbols, and are in good quantitative as well as qualitative agreement with $K_\mathrm{eff} t_\mathrm{FL}$ results from fitting to the DWMR model (full symbols in the same figure).

\begin{figure}
	\includegraphics [scale = .44, bb = 10 10 550 600]{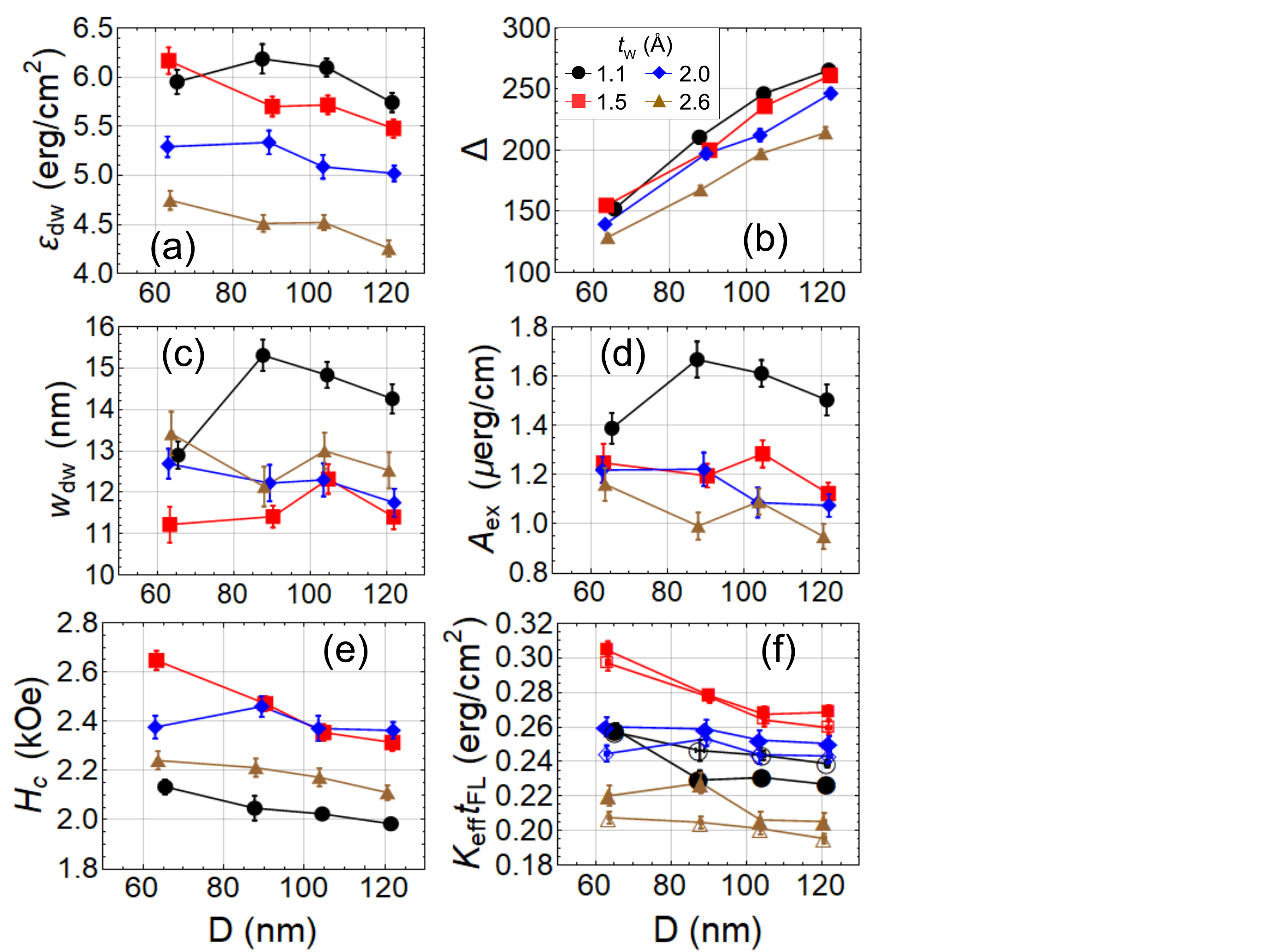}
	\caption{(a) $\varepsilon_\mathrm{dw}$, (b) $w_\mathrm{dw}$, (c) $\Delta$, (d) $A_\mathrm{ex}$, (e) $H_c$ and (f) $K_\mathrm{eff} t_\mathrm{FL} = (\ln2 / 2) (\varepsilon_\mathrm{dw} / w_\mathrm{dw}) t_\mathrm{FL}$ (full symbols) and $K_\mathrm{eff} t_\mathrm{FL} = M_s H_c t_\mathrm{FL}/ 2$ (open symbols) as a function of device diameter for FLs with various $t_\mathrm{W}$ determined by fitting $P(H)$  at  $T = 30$~$^\mathrm{o}$C using DWMR model. Each data point corresponds to median and standard error from tens of studied MRAM cells of the same nominal size. The legend shown in part (b) of the Figure refers to other parts as well.}
	\label{fig3:Edw and wdw at T30C}
\end{figure} 

This finding suggests that $\Delta$ in devices, in the DWMR regime, is determined, in addition to the $M_s$ and $A_\mathrm{ex}$ of the FL film, by the device level $H_c$, which may, or may not be directly proportional to $H_k$ of the film (compare $H_c$ values in Fig.~3(e) and $H_k$ values in Table~I). Our finding contrasts previous reports, where $K_\mathrm{eff} t_\mathrm{FL}$ measured at the film level was used to either evaluate $A_{ex}$ from  $\Delta$ measured on cells \cite{IwataHarms_SciRep2018}, or to estimate $\Delta$ from  $K_\mathrm{eff} t_\mathrm{FL}$ and $A_\mathrm{ex}$ measured on FL films \cite{IwataHarms_SciRep2019}. Lower than expected values of $A_\mathrm{ex}$ in the former case and $\Delta$ in the latter case both can be explained by our finding that $\Delta$ is more directly determined by cell-level  $H_c$ rather than  film-level $H_k$ of the FL.

We further perform the same study at higher $T = 85$~$^\mathrm{o}$C and 125~$^\mathrm{o}$C. Fig.~4 summarizes our $T$-dependent results. While $M_s$, $\varepsilon_\mathrm{dw}$, $\Delta$ and $A_\mathrm{ex}$ decrease monotonically with increasing $t_\mathrm{W}$ at all $T$s (see Figs.~4(a), 4(b), 4(d), and 4(e), respectively), $w_\mathrm{dw}$ (see Fig.~4(c)) and $K_\mathrm{eff} t_\mathrm{FL}$ (see Fig.~4(f)), show non-monotonic dependence on $t_\mathrm{W}$. In addition, one can see that $K_\mathrm{eff} t_\mathrm{FL}$ values calculated from the obtained $\varepsilon_\mathrm{dw}$ and $w_\mathrm{dw}$ (filled symbols) agree well with those obtained from $M_s H_c$ product (open symbols) for all $T$s, although for higher $T$s the latter values tend to be smaller. This discrepancy increases with increasing $T$. Our numerical modeling shows that this behavior is expected, as the ratio $ \xi = M_s H_c / (\varepsilon_\mathrm{dw} \ln2/ w_\mathrm{dw})$ has no universal value, but is a monotonically increasing function of $H_c$: $\xi \cong 1$ for $H_c = 2 - 2.5 $~kOe (corresponding to values in our experiments at $T = 30$~$^\mathrm{o}$C, see Fig.~3(e)), but decreasing for lower values $H_c < 2$~kOe that we measure at higher $T$s (see Section III and Figs~5 and 6 of the Supplementary Material).

\begin{figure}
	\includegraphics [scale = .45, bb = 10 10 550 600]{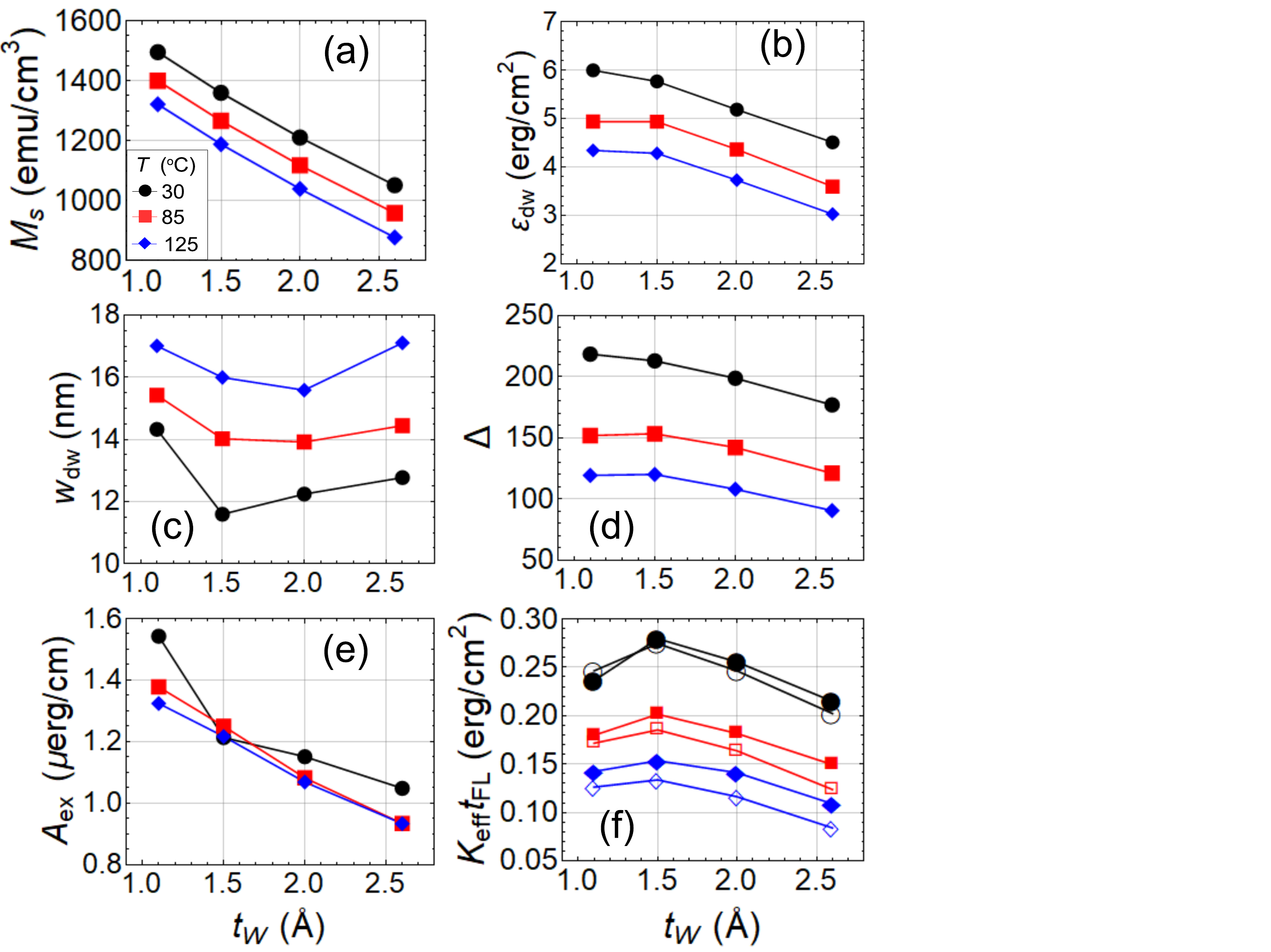}
	\caption{(a) $M_s$, (b) $\varepsilon_\mathrm{dw}$,  (c) $w_\mathrm{dw}$, (d) $\Delta$, (e) $A_\mathrm{ex}$ and (f) $K_\mathrm{eff} t_\mathrm{FL} = (\ln2 / 2) (\varepsilon_\mathrm{dw} / w_\mathrm{dw}) t_\mathrm{FL}$ (full symbols) and $K_\mathrm{eff} t_\mathrm{FL} = M_s H_c t_\mathrm{FL}/ 2$ (open symbols) as a function of thickness of W insertion layers for $T$~=~30, 85 and 125 ~$^\mathrm{o}$C. The legend shown in part (a) of the Figure refers to other parts as well. Each data point in (b)-(e) is an average value over CD = 65, 90, 105 and 120 nm.}
	\label{fig4:T dependence}
\end{figure} 

Our results show that thinner W IL could be advantageous for producing high thermal stability $\Delta$, despite having weaker PMA energy density $K_\mathrm{eff} t_\mathrm{FL}$. The advantage stems from higher $M_s$ and $A_{ex}$ of the FL film, as well as weaker $T$ dependence of the latter parameters. Our results agree with film level studies presented in References~\cite{Mohammadi_ACSAEM2019, Meo_SciRep2017, IwataHarms_SciRep2019}.

In conclusion, we report an analytical model for calculating energy barrier for domain wall mediated magnetization reversal of the MRAM cell which describes correction to the droplet model previously used in literature. Using our model, we study thermal stability factor $\Delta$ for various thicknesses of W layers inserted into FL as a function of device size and temperature. We find that,  by increasing W thickness, the effective PMA energy density of the FL film monotonically increases, but at the same time, $\Delta$ of the cell decreases. Our analysis shows that in order to maximize $\Delta$ for DWMR, one has to maximize  $M_s$ and $A_\mathrm{ex}$ of the FL film,  and $H_c$ of the cell FL. Our results also show that thinner W IL could be advantageous for producing high  $\Delta$ for MRAM cell sizes down to $\approx$~25~nm.

\bibliographystyle{aipnum4-1}

\begin{figure}
	\includegraphics [scale = 1, bb = 50 50 700 750] {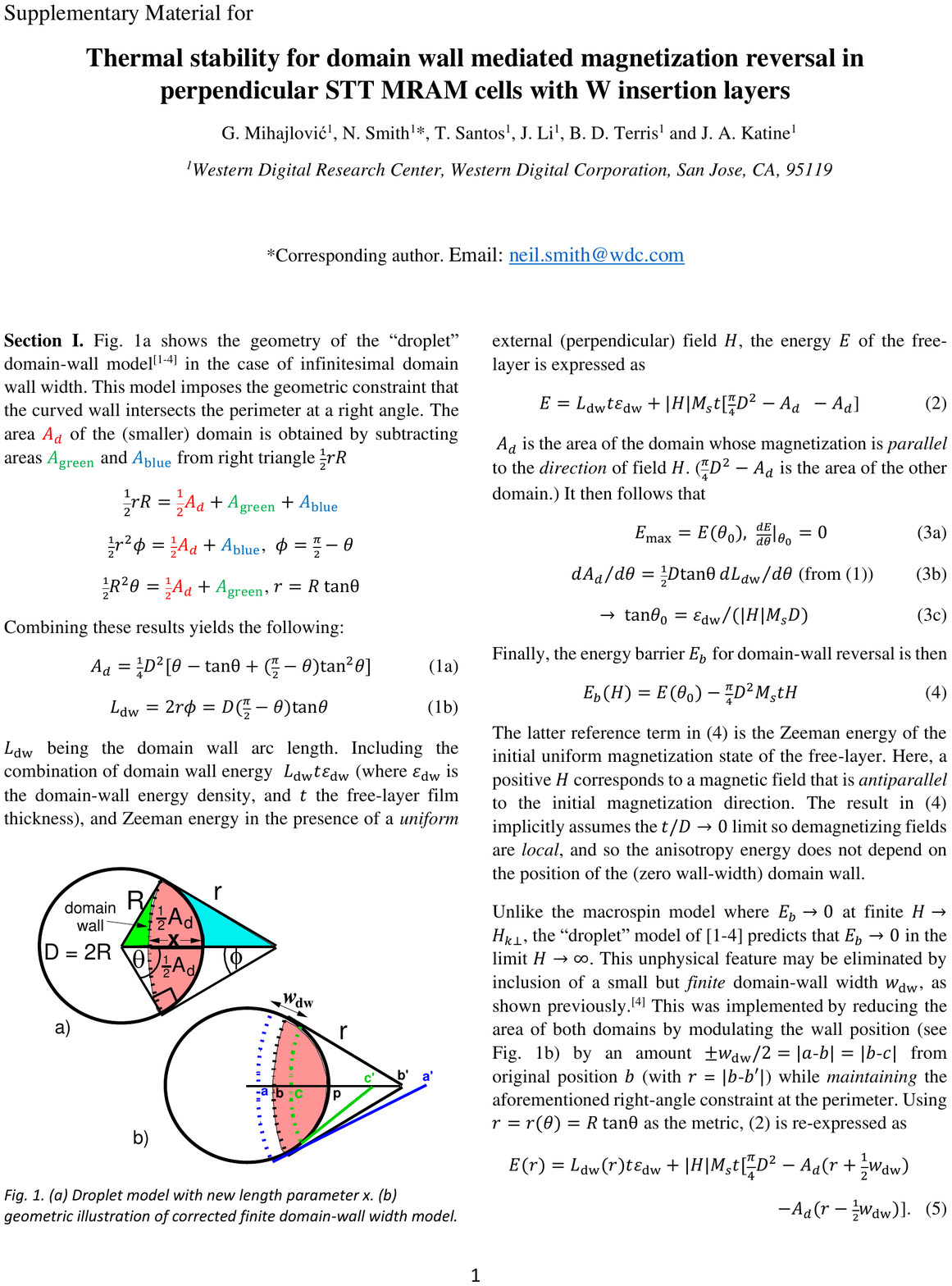}
	\label{Supplement: page1}
\end{figure}

\begin{figure}
	\includegraphics [scale = 1, bb = 50 50 700 750] {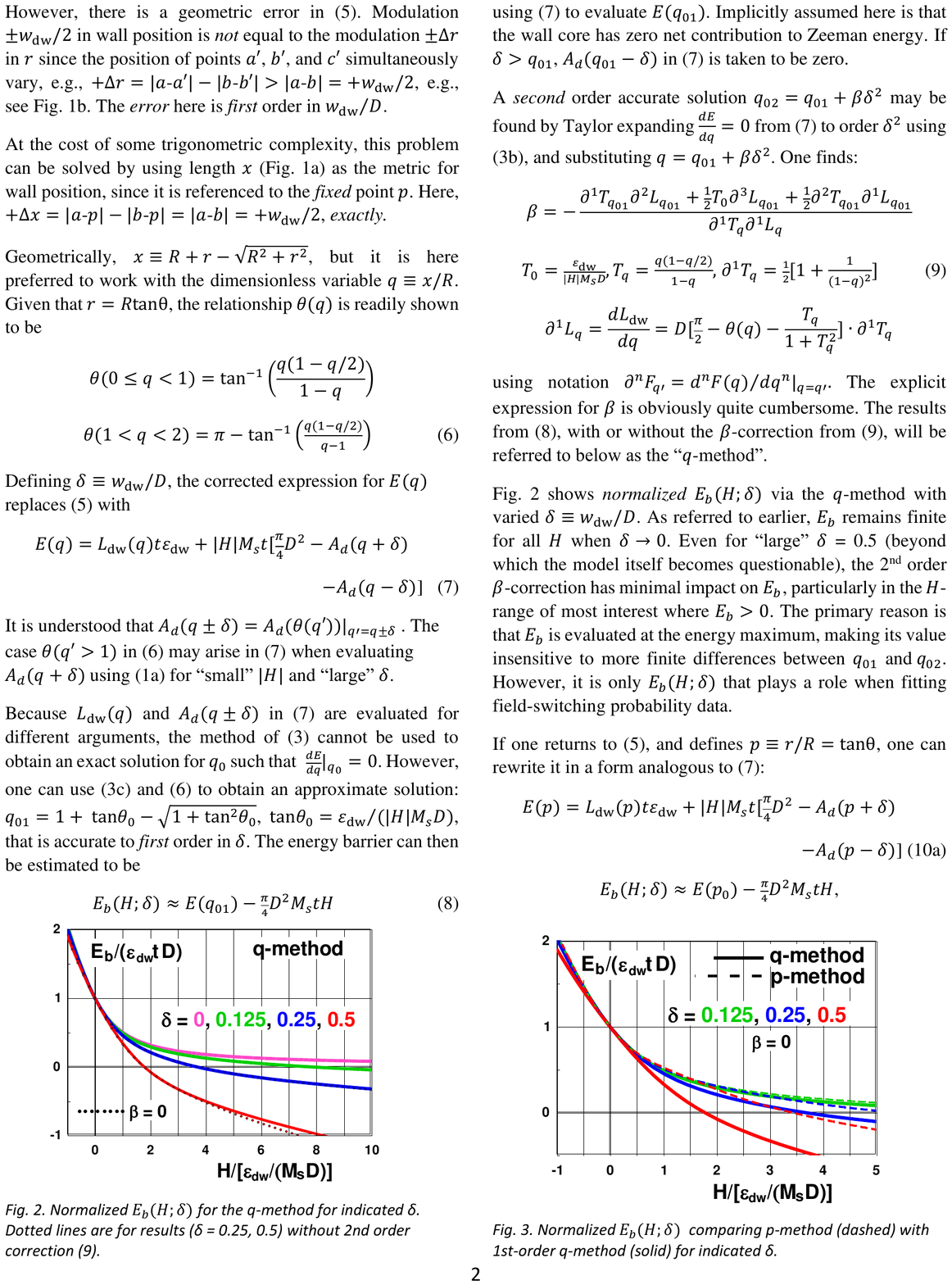}
	\label{Supplement: page2}
\end{figure} 

\begin{figure}
	\includegraphics [scale = 1, bb = 50 50 700 750] {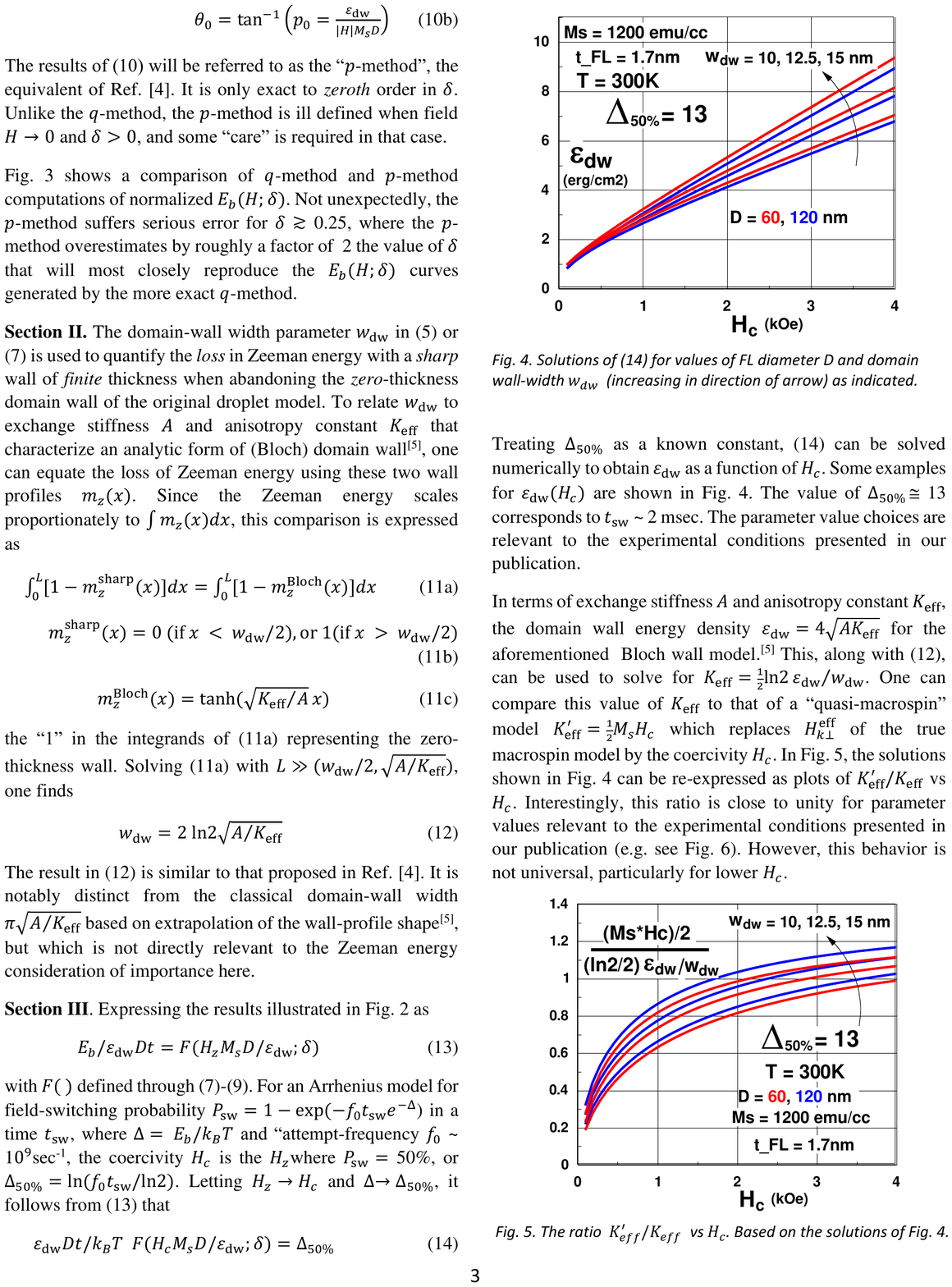}
	\label{Supplement: page3}
\end{figure} 

\begin{figure}
	\includegraphics [scale = 1, bb = 50 50 700 750] {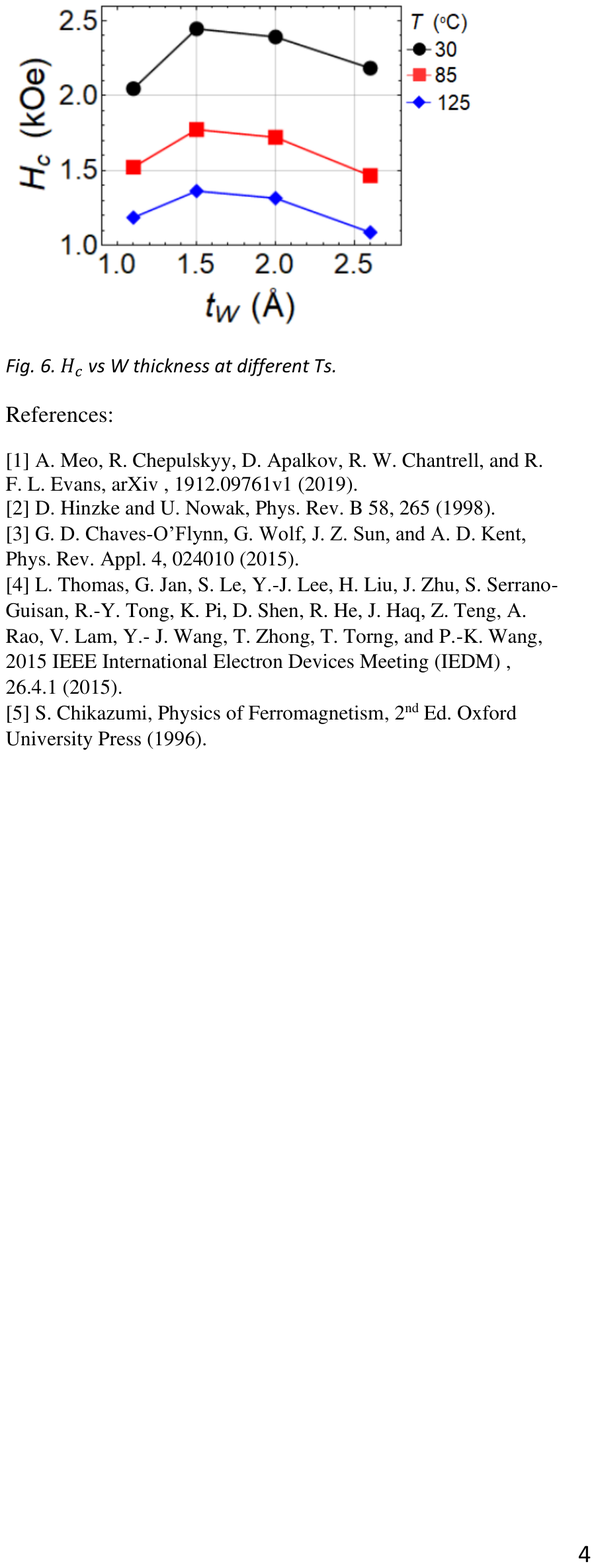}
	\label{Supplement: page4}
\end{figure} 

\end{document}